\documentclass[%
twocolumn,
superscriptaddress,
 amsmath,amssymb,
 aps,
prl,
floatfix,
letterpaper,
]{revtex4-1}

\usepackage{amsmath,amssymb,amscd,amsxtra}

\usepackage{graphicx}
\usepackage{epstopdf}

\usepackage{color}

\usepackage{subfig}
\usepackage{caption}
\newcommand\Rey{\mbox{\textit{Re}}}
\newcommand\bx{\mathbf{x}}

\newcommand\bu{\mathbf{u}}
\newcommand\ubar{\bar{\mathbf{u}}}
\newcommand\bv{\mathbf{v}}
\newcommand\bw{\mathbf{w}}

\newcommand\bU{\mathbf{U}}
\newcommand\bigR{\mathbf{R}}
\newcommand\br{\mathbf{r}}
\newcommand\be{\mathbf{e}}

\newcommand\gy{\gamma_y}            
\newcommand\dyy{\delta_{yy}}        
\newcommand\etal{\mbox{\textit{et al. }}}

\newcommand\ie{i.e.\ }

\begin{document}

\title{Pairwise interactions in inertially-driven one-dimensional microfluidic crystals}

\author{Kaitlyn Hood}
 \email{kaitlyn.t.hood@gmail.com}
\affiliation{%
 Department of Mechanical Engineering,
 Massachusetts Institute of Technology,
 Cambridge, MA 02139, USA
}%

\author{Marcus Roper}
\affiliation{%
 Department of Mathematics,
 University of California Los Angeles,
 Los Angeles, CA 90095, USA
}%

\date{\today}

\begin{abstract}
      In microfluidic devices, inertia drives particles to focus on a finite number of inertial focusing streamlines. Particles on the same streamline interact to form one-dimensional microfluidic crystals (or ``particle trains"). Here we develop an asymptotic theory to describe the pairwise interactions underlying the formation of a 1D crystal. Surprisingly, we show that particles assemble into stable equilibria, analogous to the motion of a damped spring. { The damping of the spring is due to inertial focusing forces, and the spring force arises from the interplay of viscous particle-particle and particle-wall interactions. The equilibrium spacing can be represented by a quadratic function in the particle size and therefore can be controlled by tuning the particle radius.}

\end{abstract}

\maketitle

In hydrodynamics, viscosity arises from collisions between the molecules of the fluid, transferring momentum from fast regions to slower regions.  As a result, viscosity resists large velocity gradients, and is often compared to frictional damping.  In contrast, fluid inertia maintains momentum and enhances velocity gradients in the flow. Heuristically, viscosity is thought to impede or dampen flow while inertia is thought to enhance it.  Here we present a counterexample to this intuition.  The geometry of the proposed system reverses the role of viscosity and inertia{\color{red},} so that viscous stresses perpetuate motion while inertial stresses dampen motion.

We consider the motion of two neutrally-buoyant particles suspended in a fluid moving through a rectangular channel.  The Reynolds number of the flow is chosen between 1 and 100, so that inertial stresses are equal to or greater than viscous stresses.  The fluid inertia causes the particles to migrate across streamlines and focus at finitely many inertial focusing streamlines  \cite{SegreSilberberg61, dicarlo2007continuous, DiCarlo09, ChoiSeoLee11}. Experiments  \cite{matas2004trains, humphry2010axial, lee2010dynamic, kahkeshani2016preferred, reece2016long} show that inertially focused particles ``crystallize" into trains with regular spacing (Figure \ref{fig:diagram}A-B).

There are two types of crystallization in rectangular microchannels for consideration: (i) cross-streamline crystals (which can be 2D or 3D) shown in Figure \ref{fig:diagram}B and (ii) same-streamline crystals (effectively 1D crystals) shown in Figure \ref{fig:diagram}A. Real particle trains are typically made up of a mixture of the two types \cite{kahkeshani2016preferred}.  Nonetheless, the two types of crystals have been explained by different mechanisms.  

In case (i), lattice Boltzman simulations  \cite{humphry2010axial} of the streamlines around a single inertially-focused particle showed the existence of two vortices on the opposite side of the channel (Figure \ref{fig:diagram}C reproduces these).  It was hypothesized that the centers of these vortices present stable focusing positions for a second particle.  A stable crystal forms with particles alternating between streamlines.

In case (ii) crystalization is assumed to occur at the balance of attractive and repulsive inter-particle forces. The repulsive forces appear to be symmetric, while the attractive forces appear to be non-symmetric, and therefore are believed to have separate origins \cite{lee2010dynamic}. Lee \etal hypothesize that the repulsive forces are not due to fluid inertia -- but rather are due to viscous interactions with the channel wall pushing the particles away from the focusing streamline. They assert that the attractive force arises from the inertial lift force pushing the particles back to their focusing streamlines and over-shooting, creating a harmonic oscillator type potential. 

While this mechanism gives a qualitative explanation of crystallization, it remains untested and generates more questions about the dynamics of train formation: What are the magnitudes of the attractive and repulsive forces? How do these forces depend on the experimental parameters? Can we predict the lattice length $\lambda$ as a function of the experimental parameters? While general trends are well documented, and numerical simulations can predict dynamics for a single device, there is no theoretical model that can predict the lattice length for a general class of devices and range of parameters. Such a theory could be used to engineer trains with a specific lattice length. { Controlling the lattice length is necessary in applications such as high-speed imaging, flow cytometry, and entrapment of live cells in droplets for tissue printing \cite{edd2008controlled,amini2017inertial}. A quantitative theory of lattice formation and equilibrium spacings would be one step towards rational design of such devices.} 

In order to develop our model, we analyze the interactions of pairs of particles confirming that pairs can form stable doublets in both cross-stream and same-streamline configurations. In the process of deriving the equilibrium spacing length between two particles, we discover that these stable equilibria behave like simple damped spring models where viscosity and inertia play unintuitive roles in the dynamics.

\section{Cross-streamline pairs}

\begin{figure}[ht]
\begin{centering}
\includegraphics[scale=.8]{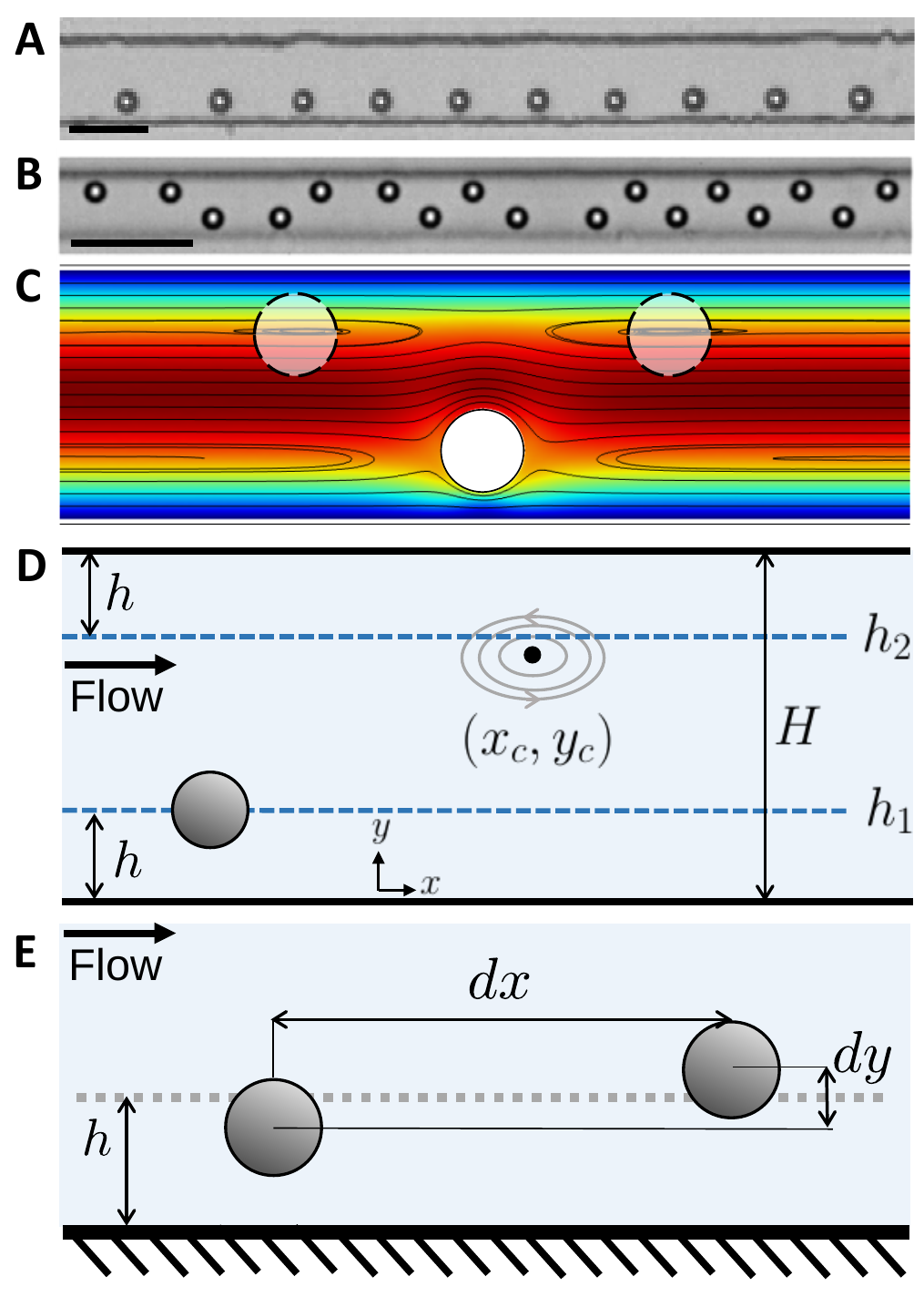}
\end{centering}
\caption{ (A) Particles on the same streamline form {a} 1D microfluidic crystal for $\Rey = 30$ and $\alpha = 0.17$ and $AR = 1.7${;} scale bar represents $90\mu$m\cite{kahkeshani2016preferred}. (B) Cross-streamline 2D microfluidic crystal{; scale bar} represents $50\mu$m\cite{lee2010dynamic}. (C) Streamlines around a single inertially-focused particle simulated using FEM in Comsol Multiphysics (Los Angeles, CA) show stagnation points on the opposite side of the channel where particles can focus to form a stable crystal with particles alternating between streamlines. { (D) Diagram for two particles focusing on opposite streamlines $h_1$ and $h_2$. The point $(x_c,y_c)$ (black dot) marks the center of the closed eddy formed by the particle focused at streamline $h_1$.} (E) Diagram for two particles near the inertial focusing streamline and a single wall.}\label{fig:diagram}
\end{figure}

First we explore the mechanism by which particles interact across streamlines.  We demonstrate mathematically how the center of a closed vortex can become a stable focusing position for a particle. Simulations \cite{humphry2010axial} of the flow around a single inertially-focused particle show closed vortices on the opposite side of the channel  (reproduced in Figure \ref{fig:diagram}C). 

{ Consider fluid flowing through a rectangular channel with height $H$, width $W$, and aspect ratio $AR = W/H$, and fluid flowing with maximum velocity $U$. If the fluid has density $\rho$ and viscosity $\mu$ then the channel Reynolds number is  $\Rey = \rho UH/\mu$. We consider two spherical particles with radius $a$ and density $\rho$ suspended in the fluid, both close to a given inertial focusing streamline.  The distance $h$ between the inertial focusing streamline and the channel wall depends on the dimensionless particle radius $\alpha = a/H$ \cite{DiCarlo09} and can be predicted from asymptotic theory \cite{Hood15}. Let $dx$ be the downstream separation of the two particles (from center to center) and $dy$ be the vertical displacement of the downstream particle above the upstream particle (Figure \ref{fig:diagram}D). }

 We assume that the original particle is on the focusing streamline $(y,z) = (h_1,0)$ and the vortices are near the focusing streamline $(y,z) = (h_2,0)$. Due to symmetry of the channel and inertial focusing, we will assume all particles are restricted to the plane $z = 0$.  Initially we treat the eddies phenomenologically; but we note that the eddies themselves can be quantitatively reproduced using the same model we develop for same streamline interactions (See the Supplemental Material).

For simplicity, we assume the closed vortex has an elliptical shape in the $x,y$-plane and is centered at $(x_c, y_c)$, where $y_c$ is sufficiently close to $h_2$. Then we can express the vortex as a second order system of ODEs: 
\begin{equation}\label{eq:eddy}
	\dot{x} = -\beta^2 (y -y_c) , \qquad
	\dot{y} = \omega^2 (x -x_c) \,.
\end{equation}
{ The direction of the eddy is determined by the location of the nearest channel wall. For example, in the case shown in Figure \ref{fig:diagram}D, because of the upper channel wall at $y = H$, the local shear flow on the streamline $y=h_2$ will be negative, \ie $-\gamma (y-h_2)$, where $\gamma > 0$. Therefore, the eddy should have a counter-clockwise orientation.}

Now we consider a second particle near the $h_2$ streamline. { We adapt the asymptotic theory developed by Hood \etal \cite{Hood15, hood2016direct} for rectangular channels. Since numerical experiments show that viscous stresses dominate momentum flux terms over the entire fluid filled domain, $V$, we can perform a regular perturbation expansion in the particle Reynolds number $\Rey_p$, treating the viscous and pressure stresses as dominant terms, and the inertial stress as a perturbative correction. 

We use the Lorentz reciprocal theorem \cite{Leal80} to represent the inertial lift force $\mathbf{F}_L$ as a volume integral that involves the following three solutions of Stokes equations ($\Rey_p = 0$): (1) $\bar{\bu}$, the undisturbed flow through the channel, (2) $\bu$, the solution for a force-free and torque-free sphere moving through the microchannel, and (3) a test velocity $\hat{\bu}$ for the slow ($\Rey_p = 0$) movement of a particle in the lateral direction in a quiescent fluid.  The total force on a particle that is constrained from migrating across streamlines can be written as an integral:
\begin{equation}
\mathbf{F}_L =  \Rey_p \int_V \hat{\bu} \cdot (\bar{\bu} \cdot \nabla \bu + \bu \cdot \nabla \bar{\bu} + \bu \cdot \nabla \bu ) \, \mathrm{dv}. \label{eq:reciprocalthm}
\end{equation}
To expose the role played by particle size in determining the lift force,we expanded $\bu$ and $\hat{\bu}$ as a two-term series in $\frac{a}{H}$, the ratio of the particle radius to the channel depth. The lift force $\mathbf{F}_L$ at the point $\bx_0$ in the channel can be expressed as a two term asymptotic expansion with coefficients $\mathbf{c}_4(\bx_0)$ and $\mathbf{c}_5(\bx_0)$.  Specifically,
\begin{equation}\label{eq:scaling_law}
      \mathbf{F}_L(\bx_0) \sim \frac{\rho U^2 a^4}{H^2} \left[ \mathbf{c}_4(\bx_0) + \frac{a}{H} \mathbf{c}_5(\bx_0)\right] .
\end{equation}
The coefficients $\mathbf{c}_4(\bx_0)$ and $\mathbf{c}_5(\bx_0)$ are dimensionless constants including both analytical and numerically computed components, and that depend on the location of the particle $\bx_0$ and the aspect ratio of the rectangular cross-section. 

To compute the inertial migration velocity in the neighborhood of $y= h_2$, we Taylor expand equation (\ref{eq:scaling_law}) around $y = h_2$. As a result, }
the particle inertial migration velocity can be expressed as  $\dot{y} = - \Gamma (y-h_2)$, where:
\begin{equation}\label{eq:gamma}
	\Gamma = \frac{ a^3 U \Rey } { 6 \pi H^4} \left(95.9 + 163.4\frac{a}{H}\right)\,.
\end{equation}

Adding inertial focusing to the system of ODEs in Eq (\ref{eq:eddy}), we arrive at:
\begin{align}\label{eq:cs_ode_dx}
	\dot{x} &= -\beta^2 (y-y_c)  , \\ \label{eq:cs_ode_dy}
	\dot{y} &= \omega^2 (x-x_c) - \Gamma (y-h_2) \,.
\end{align}
This system of ODEs has an equilibrium solution at $(x_*,y_*)$ where:
\begin{equation}
x_* = x_c + \frac{\Gamma}{\omega^2}(y_c - h_2)\,, \qquad y_* = y_c\,.
\end{equation}
We make the change of variables $X = x-x_*$ and $Y=y-y_*$, then by substitution we can re-write this as a second-order ODE in $Y$:
\begin{equation}\label{eq:cs_ode}
\ddot{Y} + \Gamma \dot{Y} + \omega^2 \beta^2 {Y}  = 0 \,.
\end{equation}
The right hand side of equation (\ref{eq:cs_ode}) equal to zero if we choose $y^*$ to be:
\begin{equation}
	y^* = \frac{\omega^2 \beta^2 y_c + \Gamma h_2}{\omega^2 \beta^2 + \Gamma} \,.
\end{equation}
Then equation (\ref{eq:cs_ode}) becomes a homogeneous second-order differential equation with constant coefficients, or a damped harmonic oscillator.  We see that the damping term is proportional to $\Gamma$, the inertial focusing constant.  As a result, the particle focuses to $(X,Y) = (0,0)$ or $(x,y) = (x_c, y^*)$.

We have shown that the inertially-driven damping of particle motion in an eddy forces the particle to focus to a single point.  Notice that the focusing position of the particle is not exactly on the inertial-focusing streamline, but at a weighted average between the streamline and the center of the eddy, where the weights are the inertial focusing constant $\Gamma$ and the elliptical eddy constants $\beta$ and $\omega$. 

This analysis provides a mechanism by which particles can form stable cross-stream pairs. However, it does not appear to apply to same-streamline crystals because there are no closed eddies on the same streamline as the focused particle, only a recirculating flow (Figure \ref{fig:diagram}C).  In order to explain same-streamline crystallization, we need to derive a new model from first principles.

\section{Same-streamline pairs}

Here we derive a model for the assembly of pairs of same-streamline crystals.  In order to make an asymptotic expansion, we assume that $ a \ll h \ll dx$.

\begin{figure}[t]
\begin{centering}
\includegraphics[scale=.7]{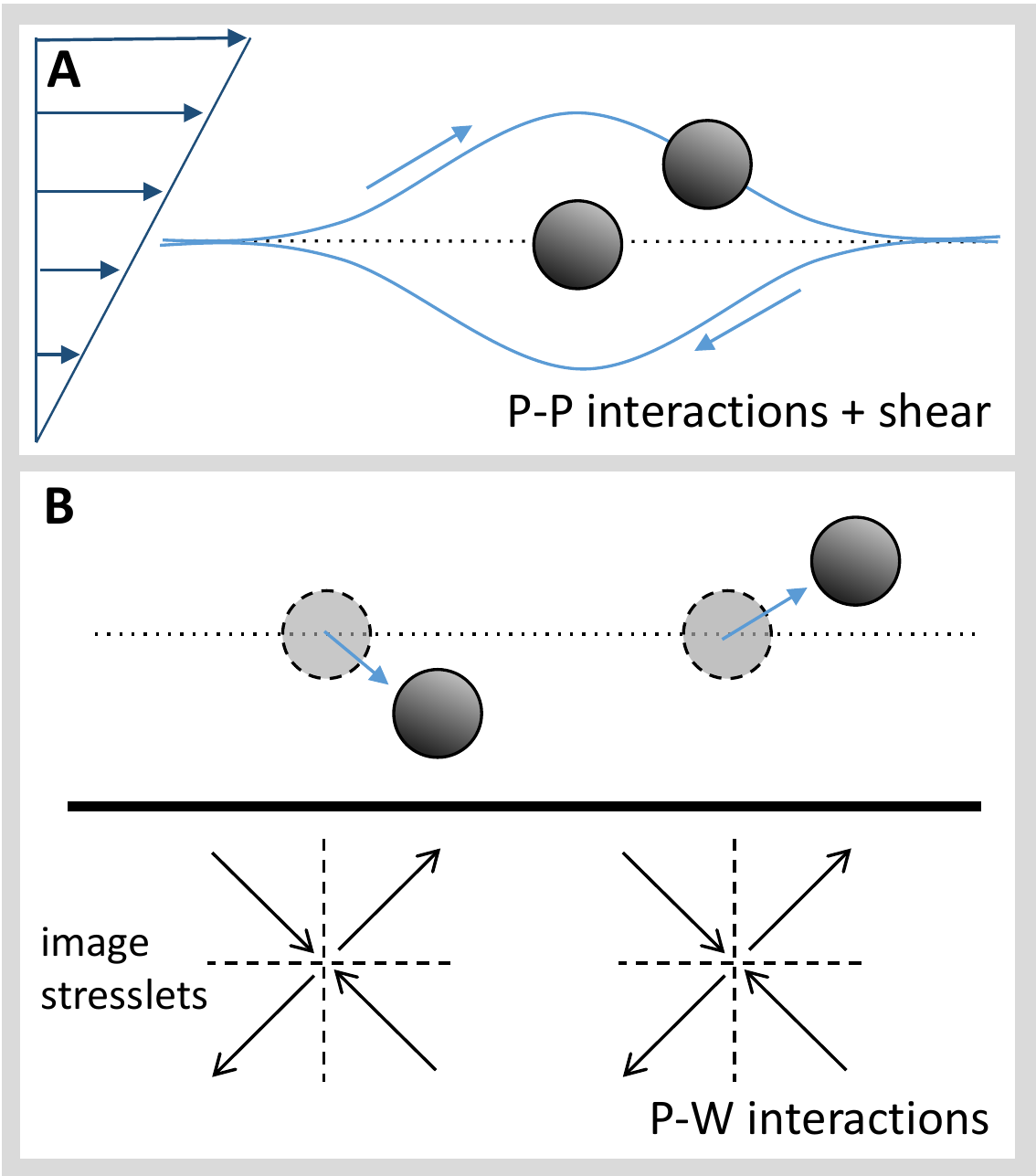}
\end{centering}
\caption{ (A) Viscous P-P interactions { in a shear flow predicts `bound' pairs of spheres with closed trajectories \cite{batchelor1972hydrodynamic} and} with $\dot{dy} < 0 $. Shown in the moving reference frame of one particle. (B) Viscous P-W interactions can be represented by image stresslets.  The image on particle 1 acts on particle 2 and vice versa, creating a net $\dot{dy} > 0$. }\label{fig:pp_pw_inter}
\end{figure} 

{  In a rectangular channel flow, numerical experiments show that viscous stresses dominate over momentum flux terms over the entire channel \cite{Hood15}}.  Hence, a three-dimensional asymptotic analysis of the Navier-Stokes equations for this system  showed that a low Reynolds number approximation is valid. This analysis demonstrated that the dominant physics is viscous, and that inertial focusing can be treated as a perturbative effect.  

What are the { essential ingredients} needed to model the interactions of a pair of particles within a same-streamline 1D crystal?  First, we need inertial focusing to constrain the particles on a streamline.  Second, we need particle-particle (P-P) interactions.   Third{\color{red},} we need the local background flow (i.e. the flow in a channel undisturbed by particles), which to first order is a shear flow.  Fourth, we find that it is necessary to include particle-wall (P-W) interactions (with the nearest channel wall) in order to achieve a stable configuration.  The role of the P-W interactions will be made clear later in this section.  

{ Because the asymptotic theory that accurately predicts the lift force in Eq (\ref{eq:reciprocalthm}) arises from a perturbation expansion in small $\Rey_p$, we conclude that, in a channel geometry, viscous effects are first order and inertial effects are second order \cite{Hood15}. Therefore, it suffices to approximate the P-P interactions and the P-W interactions with their viscous counterparts.}  Furthermore, these viscous interactions can be written analytically as a multipole expansion \cite{batchelor1972hydrodynamic, dacunha1996shear}.  Likewise, inertial focusing can be written as a two-term asymptotic series whose coefficients were computed numerically by Hood \etal \cite{Hood15}.

Viscous P-P interactions in a shear { flow results in `bound' pairs of spheres with closed trajectories \cite{batchelor1972hydrodynamic}} (Figure \ref{fig:pp_pw_inter}A). { We will re-derive this result using Lamb's solution, the method of reflections, and Fax{\'e}n's laws in Section \ref{sec:model} and add additional physics.} Because this orbit is clockwise in the sense of the coordinates used in Figure \ref{fig:diagram}D and \ref{fig:pp_pw_inter}A, {and because we have defined $dy$ to be the vertical displacement between the leading and trailing particle, we observe that $\dot{dy}$ is negative throughout. Starting with the two spheres with $dx \sim 0$, then $dy$ is positive. As the particles orbit, $dy$ decreases monotonically and passes through zero and then becomes negative. The vertical displacement $dy$ reaches its minimum value when $dx = 0$, at which point the trailing particle becomes the leading particle. During this first phase of the orbit, $\dot{dy}$ was negative throughout. In the second phase, after the leading and trailing particles switch, $dy$ starts out positive and decreases monotonically to a negative value, resulting in a negative $\dot{dy}$.}

Viscous P-W interactions act in the opposite direction on the vertical displacement $dy$.  We can see this by using the method of images to model the effect of the wall on the particles.  To first order, we approximate the image particles by stresslets.  The induced velocity on the downstream particle is calculated by evaluating the upstream image stresslet at the center of the downstream particle and has a positive $y$ component.  Likewise the induced velocity on the upstream component has a negative $y$ component, so that the net vertical displacement $dy$ is positive (Figure \ref{fig:pp_pw_inter}B).

\begin{figure}[t]
\begin{centering}
\includegraphics[scale=.34]{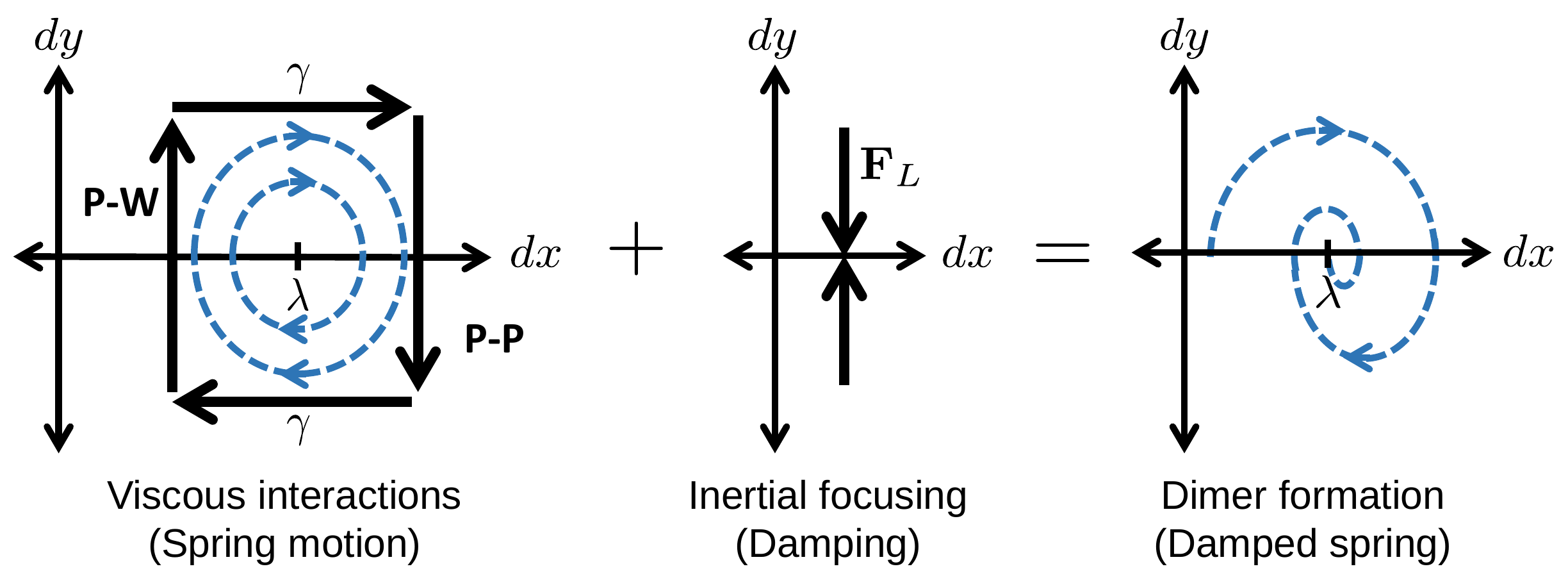}
\end{centering}
\caption{ Analogy between nucleation and damped spring motion. }\label{fig:spring_diagram}
\end{figure}

{ The shear flow centered at the height $h$ converts any vertical displacement $dy$ into a streamwise displacement $dx$.  Combining the shear flow with viscous P-P interactions and viscous P-W interactions creates a closed loop with an equilibrium point at $(dx, dy) = (\lambda, 0)$ (Figure \ref{fig:spring_diagram} left).  In dynamical systems, $(\lambda, 0)$ is called a center and is neutrally stable. Note that when the particles are on the same streamline, neither P-P nor P-W interactions act to alter the spacing $dx$ directly.  The equilibrium shows up as a point where $\dot{dy}$ vanishes.  Thus, it is not detected using the standard approach to finding equilibria (i.e. analyzing where $\dot{dx}=0$). }

In contrast, inertial focusing acts uniformly on particles, regardless of their separation $dx$, and always pushes particles back to the inertial focusing streamline at $y=h$.  Therefore inertial focusing pushes $dy$ to zero (Figure \ref{fig:spring_diagram} center).  Adding inertial focusing to the viscous system above creates an asymptotically stable spiral point that converges to $(dx, dy) = (\lambda, 0)$ (Figure \ref{fig:spring_diagram} right).

\begin{figure*}[t]
\begin{centering}
\includegraphics[scale=.85]{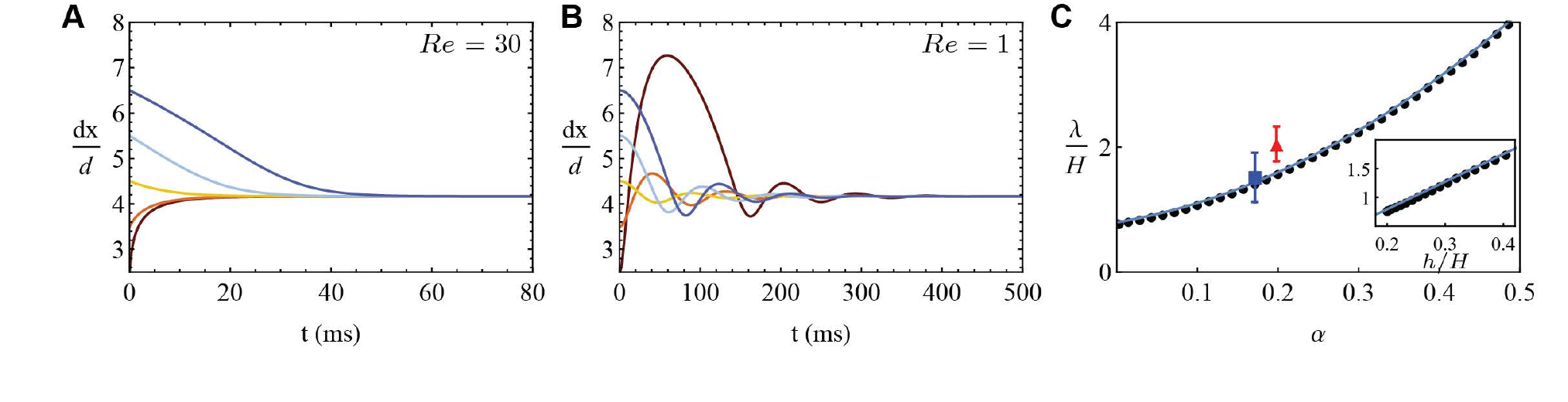}
\end{centering}
\caption{ The separation of two particles $dx(t)$ as a function of time for $a = 6\mu$m,  different initial separation lengths, and (A) $\Rey = 30$ or (B) $\Rey = 1$.  (C) The equilibrium separation length $\lambda$ is a function of the relative particle size { $\alpha = a/H$} { and equation (\ref{eq:2p_lambda}) captures this realationship well}.  Here the markers represent numerical solutions to equations (\ref{eq:ode_dx})-(\ref{eq:ode_dy}) and the solid line is equation (\ref{eq:2p_lambda}). Experimental measurements from Kahkeshani \etal \cite{kahkeshani2016preferred} at $\Rey_p = 2.8$ (blue square) and Lee \etal \cite{lee2010dynamic} (red triangle) agree with our model.  Error bars are standard deviations. (Inset) We observe that $\lambda$ is a linear function of $h$ and can be approximated by equation (\ref{eq:lambda_H}).}\label{fig:ode}
\end{figure*}

The dynamics of the system of two inertially-focused particles interacting mimics the behavior of a damped harmonic oscillator or a spring with frictional damping (Figure \ref{fig:spring_diagram}).  Here the viscous interactions are analogous to the spring motion creating closed trajectories in $(dx,dy)$ space while inertial focusing is analogous to frictional damping.  Herein lies the role-reversal: viscosity maintains motion (like a spring) and inertia dampens motion (like friction).

\section{Dynamic model of crystallization}\label{sec:model}

We can make this description rigorous by writing down the equations of motion and solving them numerically. Let $\bx_i$ for $i=1,2$ be the locations of the two particles. { We begin by finding the exact solution for the flow around an unbounded parabolic flow around a single force-free and torque-free no-slip sphere.} The flow around each particle can be derived using {Lamb's} solution for the flow exterior to a sphere \cite{Lamb45,KimKarrila2005}. Here we will only keep the terms that are $O(r^{-2})$ and $O(r^{-3})$.  In order to derive the image system in the next step, we must convert {Lamb's} solution into multipole singularities.  In this case the $O(r^{-2})$ term becomes the stresslet $\bv^{\mathrm{ST}}$, and the $O(r^{-3})$ term is decomposed into the source dipole $\bv^{\mathrm{D}}$ and two stokeslet quadrupoles $\bv^{\mathrm{SQ}}$ and $\bw^{\mathrm{SQ}}$.  See Supplemental material at (link) for detailed derivation of these terms and their images below. 

For each particle, we model the viscous wall effects by computing the image system for a plane wall.   Blake \cite{blake1971note} derived the image system for a stokeslet, and using a similar procedure the image systems for the stresslet $\bv^{\mathrm{STim}}$, source dipole $\bv^{\mathrm{Dim}}$, and stokeslet quadrupoles $\bv^{\mathrm{SQim}}$ and $\bw^{\mathrm{SQim}}$ can be derived \cite{blake1974fundamental, spagnolie2012hydrodynamics}. Then the flow around each particle is: 
\begin{align}\label{eq:vel_1p}
      \bv_i &\sim \,  (\bv_i^{\mathrm{ST}} + \bv_i^{\mathrm{STim}}) + (\bv_i^{\mathrm{D}}  + \bv_i^{\mathrm{Dim}}) \\ &+ (\bv_i^{\mathrm{SQ}} + \bv_i^{\mathrm{SQim}}) + (\bw_i^{\mathrm{SQ}} + \bw_i^{\mathrm{SQim}}) \,. \nonumber
\end{align}
{Corrections to $v_i$ from the presence of particle $j\neq i$ are higher order and therefore not included in this step.}

Let $\ubar$ be the Poiseuille flow through a rectangular channel \cite{Papanastasiou99}. Then, for each particle we { use Fax{\'e}n's law \cite{KimKarrila2005} to} compute the induced velocity from the other particle and image system,
\begin{align}
      \bU_i &= \left( 1 + \frac{a^2}{6} \nabla^2 \right) (\ubar + \bv_j) \Big|_{\bx = \bx_i} \, , \qquad i \neq j.
\end{align}
{ Here, $\bU_i = (U_i,V_i,W_i)$.} Then we define the relative velocity $d\bU = \bU_2 - \bU_1$.  

We again model inertial focusing by Taylor expanding the migration velocity from Hood \etal \cite{hood2016direct} in the coordinate $y$ around $h$.  This gives $\dot{y}_i = - \Gamma (y_i - h)$, where the inertial focusing constant $\Gamma$ is defined in equation (\ref{eq:gamma}). Combining the viscous particle interactions $d\bU = (dU,dV,0)$ with the inertial focusing we arrive at a system of ODEs for the dynamics of particle interactions:
\begin{align}\label{eq:ode_dx}
      \dot{dx} &= dU, \, \qquad \qquad \qquad   dx(t=0) = k_0d, \\ \label{eq:ode_dy}
      \dot{y_i} &= V_i - \Gamma (y_i - h) , \quad y_i(t=0) = h, \quad i=1,2. 
\end{align}
The ODEs depend explicitly on the particle size $\alpha$, the Reynolds number $\Rey$, and the initial separation length $k_0 d$.  The equations implicitly depend on the channel aspect ratio $AR$, but throughout this paper we will consider the same channel as Kahkeshani \etal \cite{kahkeshani2016preferred}, where $W = 60\mu$m, $H = 35\mu$m, and $AR = 1.7$. 

Solving ODEs (\ref{eq:ode_dx})-(\ref{eq:ode_dy}) numerically for $\Rey = 30$, $a = 6\mu$m, and various initial conditions shows that there is a stable equilibrium length $\lambda = 4.17d$ (Figure \ref{fig:ode}A).  In contrast, the same system for $\Rey = 1$ converges to the same value of $\lambda = 4.17d$, but the harmonic oscillator becomes under-damped (Figure \ref{fig:ode}B).  This shows that as $\Rey$ increases, so does the damping of the spring motion. { This behavior is counters the intuition that viscosity should play the damping role, not the inertia.}

How does the lattice length $\lambda$ scale with experimental parameters?  Contrary to expectations, we find that $\lambda$ does not scale linearly with particle diameter $d = 2a$. From the derivation of our asymptotic model, we would expect $\lambda$ to depend on both the particle radius $a$ and the distance from the inertial-focusing streamline to the wall $h$. Surprisingly, we find from the numerical solutions of equations (\ref{eq:ode_dx})-(\ref{eq:ode_dy}) that $\lambda$ depends linearly on $h$ (Figure \ref{fig:ode}C Inset).  A polynomial fit of the numerical data predicts that:
\begin{equation}\label{eq:lambda_H}
      \lambda = -0.2H + 4.8h \,.
\end{equation}
We conjecture that $h$ is the scaling parameter for the equilibrium spacing, instead of $\lambda$.  It is not suprising that $h$ influences $\lambda$ strongly because $h$ appears in the P-W interaction term, which was necessary to include in our model in order to form stable equilibria.  In terms of the qualitative model of Lee \etal \cite{lee2010dynamic}, the P-P interactions give rise to a repulsive force between the particles while the P-W interactions lead to an attractive force.  Since the strength of the P-W interactions depend explicitly on $h$, it follows that $h$ should strongly determine the equilibrium spacing $\lambda$.

Additionally, $h$ depends implicitly on the relative particle size $\alpha = a/H$ (recall that $H$ is the height of the channel), and can be approximated by a quadratic polynomial \cite{Hood15}.  Therefore, we expect that $\lambda$ can be expressed as a function of the relative particle size $\alpha$.  Using a similar analysis, we observe that for infinitesimal particle sizes, the equilibrium spacing $\lambda$ approaches a constant $\lambda \sim 0.8H$ (Figure \ref{fig:ode}C). As particle size $\alpha$ increases, $\lambda$ also increases.  A polynomial fit of the numerical data for $\lambda$ predicts that:
\begin{equation}\label{eq:2p_lambda}
      \frac{\lambda}{H} = 0.8 + 2.2\alpha +9.1\alpha^2 \,.
\end{equation}
We compare the numerical data and the numerical fit in equation (\ref{eq:2p_lambda}) to experimental data from Kahkeshani \etal \cite{kahkeshani2016preferred} (at $\Rey_p = 2.8$) and Lee \etal \cite{lee2010dynamic}. { Our model with no fitting parameters (\ref{eq:2p_lambda}) matches well with the experimental data (Figure \ref{fig:ode}C). This fit persists even though the channels have different aspect ratios ($AR = 1.7$ and $AR = 3.6$, respectively), suggesting that the modeling assumption that the flow is predominantly 2-D is valid.} 

We note that the equilibrium spacing $\lambda$ is independent of $\Rey$ in our theory (though our theory is asymptotically correct as $\Rey \to 0$, so higher order corrections are needed to model the effect of $\Rey$ on the equilibrium spacing).  {In our model}, $\Rey$ does not impact the equilibrium of the system, only the degree of damping.

\section{Crystallization at moderate Reynolds numbers}

In our model, we assume particle interactions are dominated by viscosity{\color{red},} which is asymptotically correct in the limit of small Reynolds numbers.  However, particle train formation still occurs at moderate $\Rey$, and preferred spacings of particles can change as $\Rey$ increases \cite{kahkeshani2016preferred}.

{  Kahkeshani \etal \cite{kahkeshani2016preferred} measured the inter-particle spacings of particle trains as the particle Reynolds number $\Rey_p$ changes. At $\Rey_p = 2.8$ they measured a pdf of particle spacings that yielded $\lambda = (4.4\pm1.2)d$, which agrees with our theoretical prediction of $\lambda = 4.17d$ in Section \ref{sec:model} (Figure \ref{fig:ode}C). However, at $\Rey_p = 8.3$, they measure $\lambda = (2.0 \pm 0.3)d$, which does not agree with our theory. While we expect that our theory is valid only at lower values of $\Rey_p$, some insight into train formation at intermediate $\Rey_p$ can be gleaned from examining particle paths.}

As a first step toward a physical theory for crystallization at moderate Reynolds numbers, we adopt an approach recently used to study particle chaining in acoustic streaming flows  \cite{klotsa2007interaction, klotsa2009chain}.  We analyze the vortical structures created by single particles and then look for patterns of interference between particles. Klotsa \etal {found} empirically that particles tend {to} organize themselves into configurations that minimize total kinetic energy in the surrounding flow \cite{klotsa2009chain}.

\begin{figure}[t]
\begin{centering}
\includegraphics[scale=1]{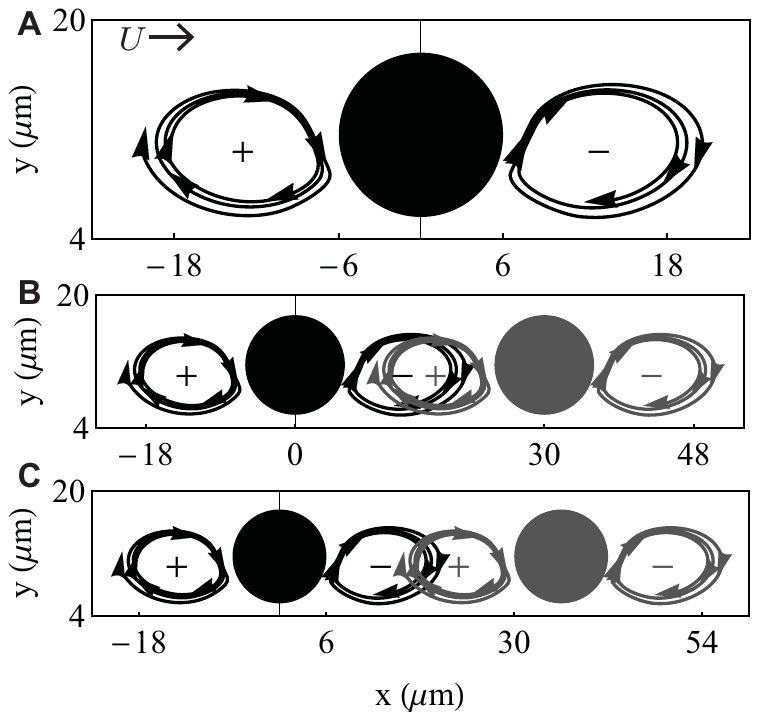}
\end{centering}
\caption{ (A) {A particle on the inertial focusing streamline acts on a second particle, drawing vortical path-lines both upstream and downstream}. The trailing vortex spirals outward while the leading vortex spirals inward. (B) If particles approach too closely, then vortices interfere constructively. (C) If particles are spaced further apart, the vortices interfere destructively.
}\label{fig:1p_streamlines}
\end{figure}

We investigate the approximate velocity around a single inertially-focused particle $\bU_i$. Note that the trajectories of $\bu_i(\bx)$ show the paths that another particle would follow if introduced at a point $\bx_0 = (x_0,y_0)$; they are therefore particle paths, not streamlines. { These particle paths are explained by the schematic in Figure \ref{fig:spring_diagram}. Specifically, we constrain particle 1 to the streamline $y=h$, then the path of particle 2 $(x(t),y(t))$ would satisfy:
\begin{align}\label{eq:paths_x}
\dot{x} &= w_x \,, \qquad x(0) = x_0,  \\ \label{eq:paths_y}
\dot{y} &= w_y \,, \qquad y(0) = y_0, 
\end{align}
where $\mathbf{w} = (w_x,w_y,w_z)$ satisfies:
\begin{equation}
\mathbf{w} = \left( 1 + \frac{a^2}{6} \nabla^2 \right) \bv_1 \bigg\rvert _{x_1 = 0,y_1 = h} \,.
\end{equation} }
{ Notice that $\mathbf{w}$ is the induced flow of particle 2 due to particle 1. It is not $\bv_1$, the flow around particle 1, which could be compared directly to the numerical simulation of the flow around an inertially focused particle (Figure \ref{fig:diagram}C).}  { In our analysis, we consider only one-way interactions, so particle 1 does not leave its inertially-focused position.  }

We observe that the particle paths form a leading vortex and a trailing vortex both with the same sense of rotation (Figure \ref{fig:1p_streamlines}A).  On closer observation we notice that neither structure is closed. These zones of recirculation have been observed experimentally \cite{kahkeshani2016preferred}.  The leading vortex is an inward spiral, while the trailing vortex is an outward spiral (Figure \ref{fig:1p_streamlines}).  Closure (or not) of the eddies is not a significant factor in our {subsequent} analysis.

\begin{figure}[t]
\begin{centering}
\includegraphics[scale=.6]{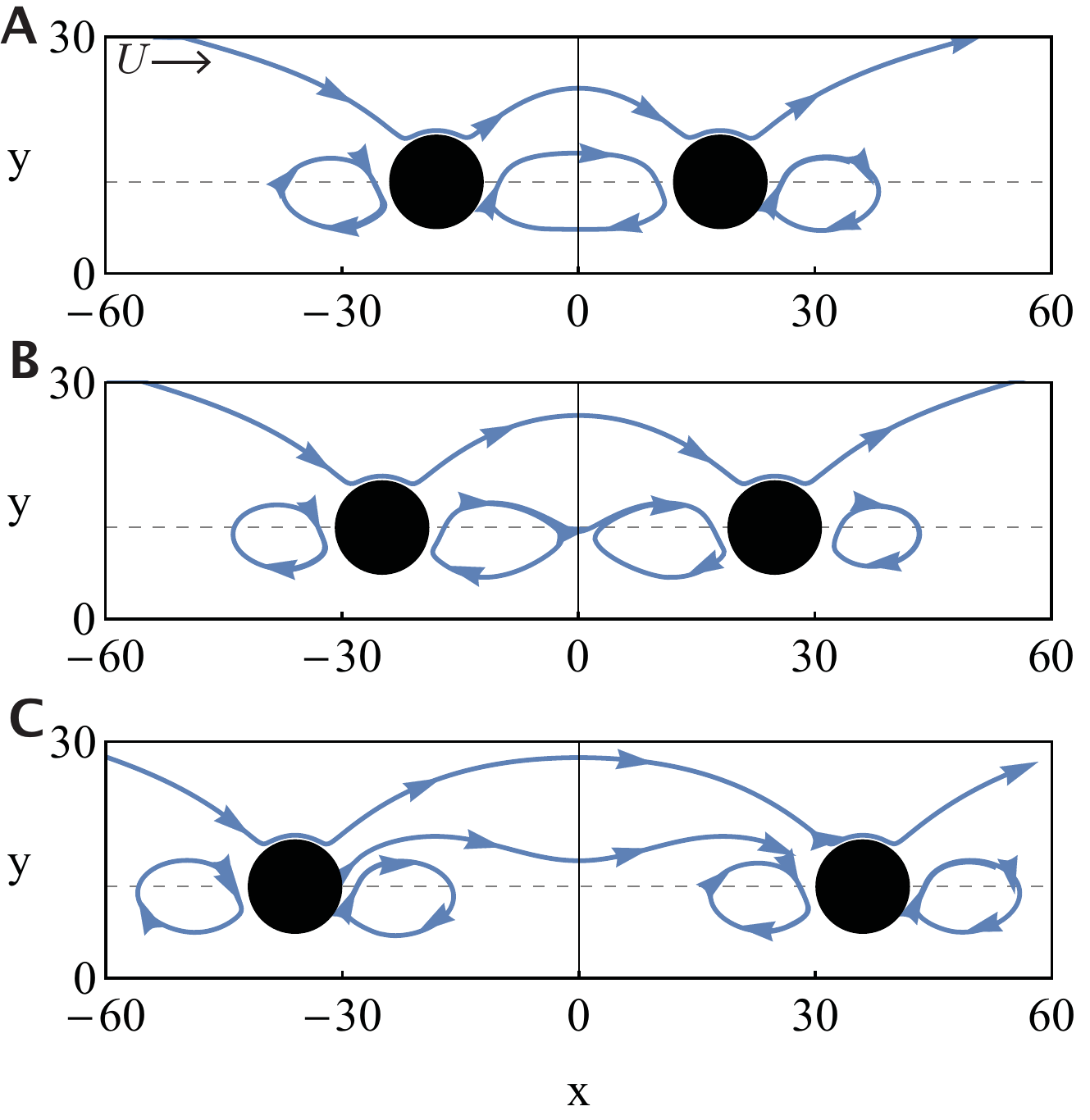}
\end{centering}
\caption{ Vortex interactions for two inertially-focused particles. (A) $dx = 3d < \lambda$, (B) $dx = 4.15d = \lambda$, (C) $dx = 6d > \lambda$. }\label{fig:2p_streamlines}
\end{figure}

{ There is an optimum spacing between the particles that minimizes total kinetic energy.}  If the particles are brought close enough together, then the vortices overlap and reinforce each other, as shown in Figure \ref{fig:1p_streamlines}B.  No longer cancelling, the kinetic energy of the flow will now increase. The orientation of the vortices agree with the pair trajectories computed in Kahkeshani \etal \cite{kahkeshani2016preferred}. { Conversely, when two particles are spaced far apart, their respective leading and trailing vortices will tend to cancel each other, as shown in the schematic in Figure \ref{fig:1p_streamlines}C. } Following the reasoning of Klotsa \etal \cite{klotsa2009chain}, we expect the particles to self-organize into a configuration that minimizes the kinetic energy, i.e. intermediate between Figures \ref{fig:1p_streamlines}B and \ref{fig:1p_streamlines}C.

We confirmed that these predictions are supported in our simulations of particles interacting at small Reynolds numbers. { We compute particle paths around two inertially-focused particles separated by a distance $\lambda$, \ie the particles are located at $(x_1,y_1) = (0,h)$ and $(x_2,y_2) = (dx,h)$. Then, the particle paths are determined by equations (\ref{eq:paths_x})-(\ref{eq:paths_y}) where
\begin{equation}\label{eq:pair_paths}
\mathbf{w} = \left( 1 + \frac{a^2}{6} \nabla^2 \right) (\bv_1 + \bv_2 )\bigg\rvert _{x_1 = 0,x_2 = dx,y_1 = y_2 = h} \,.
\end{equation}

The paths determined by equations (\ref{eq:paths_x})-(\ref{eq:paths_y}) and (\ref{eq:pair_paths}) represent the interference of the vortices in Figure \ref{fig:1p_streamlines}B-C. } When the two particles are close together, $dx < \lambda$, then the two vortices combine to form a closed ring (Figure \ref{fig:2p_streamlines}A). { The two vortices overlap and reinforce each other, thereby increasing the total kinetic energy of the system.} When the particles are too far apart $dx > \lambda$, the {vortices} cancel only {weakly} (Figure \ref{fig:2p_streamlines}C). At the center point of the particles, the paths are clearly unstable. Conversely, when the particles are at their equilibrium spacing $dx = \lambda$, the vortices connect to each other but maintain their distinct centers (Figure \ref{fig:2p_streamlines}B). { In this configuration, the vortices cancel at the midpoint creating a third stagnation point, which decreases the total kinetic energy. }

{ As $\Rey_p$ increases, we expect that the boundary layers on the particles should decrease. According to Kahkeshani \etal \cite{kahkeshani2016preferred}, we would expect that, at some critical $\Rey_p$, a new pair of vortices appear closer to the particle in Figure \ref{fig:1p_streamlines}A. Since the size and location of the vortices determine the equilibrium spacing $\lambda$ between the particles, we would expect that higher $\Rey_p$ particle trains should have smaller $\lambda$.}

\section{conclusions}

Under our model, pairs of particles organize into stable equilibria that are analogous to damped springs, in which the expected roles of inertia and viscosity have been reversed. Viscous flow maintains harmonic motion, like a spring, while inertial focusing results in a damping effect. 

 The { essential ingredients} needed to model the harmonic motion are: shear flow, particle-particle interactions and particle-wall interaction{s}. We showed that particle-wall interactions are necessary to achieve negative vertical displacement $dy$, and therefore necessary to achieve closed trajectories in the viscous harmonic motion.

We developed an asymptotic model to describe this behavior and produced a formula for the lattice spacing $\lambda$. We envisage that the model for particle spacing (the terms of which are directly written out in the Supplementary Material) will be generally useful for reduced order simulations for particles in inertial microfluidic devices.  We showed that $\lambda$ scales with the distance $h$ between the inertial focusing streamline and the channel wall.  Since the distance $h$ depends on the relative particle size $\alpha = a/H$, the lattice spacing $\lambda$ can be tuned by changing particle sizes.  As a result, not only is the effect of the channel walls necessary to model the dynamics, but it also sets the scaling for the lattice length.

Additionally, we have shown that both the cross-stream pairs
and same-stream pairs form a stable configuration when a closed particle path is combined with inertial focusing to a streamline. In the case of same-stream pairs, the closed particle path is not apparent at the level of the fluid velocity, and requires asymptotic approximations to reveal the underlying vortical structure of the system.

\section{Acknowledgments}

This material is based upon work supported by the National Science Foundation under Award No. DMS-1606487 (to K.H.) and DMS-1312543 (to M.R.).  This work was partially supported by the UCLA Dissertation Year Fellowship (to K.H.). We thank Lawrence Liu for performing preliminary simulations (supported by DMS-1045536), and Hamed Haddadi and Soroush Kahkeshani for helpful discussions.

 
\providecommand{\noopsort}[1]{}\providecommand{\singleletter}[1]{#1}%
%

\widetext
\clearpage
\begin{center}
\textbf{\large Supplemental Materials: Nucleation of inertially-driven one-dimensional microfluidic crystals}\\
Kaitlyn Hood and Marcus Roper
\end{center}
\setcounter{equation}{0}
\setcounter{figure}{0}
\setcounter{table}{0}
\setcounter{page}{1}
\makeatletter
\renewcommand{\theequation}{S\arabic{equation}}
\renewcommand{\thefigure}{S\arabic{figure}}
\renewcommand{\thesection}{S\arabic{section}}
\renewcommand{\bibnumfmt}[1]{[S#1]}
\renewcommand{\citenumfont}[1]{S#1}

\section{Derivation of the viscous model for two particles}

Here we derive an asymptotic model of two particles in a Poiseuille flow near a wall.  We use this model to predict the equilibrium spacing of two particles in an inertial microfluidic device.  We compare our asymptotic model to numerical simulations of the full Navier-Stokes equation, themselves fully validated in section \ref{sec:validate}.

\begin{figure}[h]
\centering
\includegraphics[scale=.4]{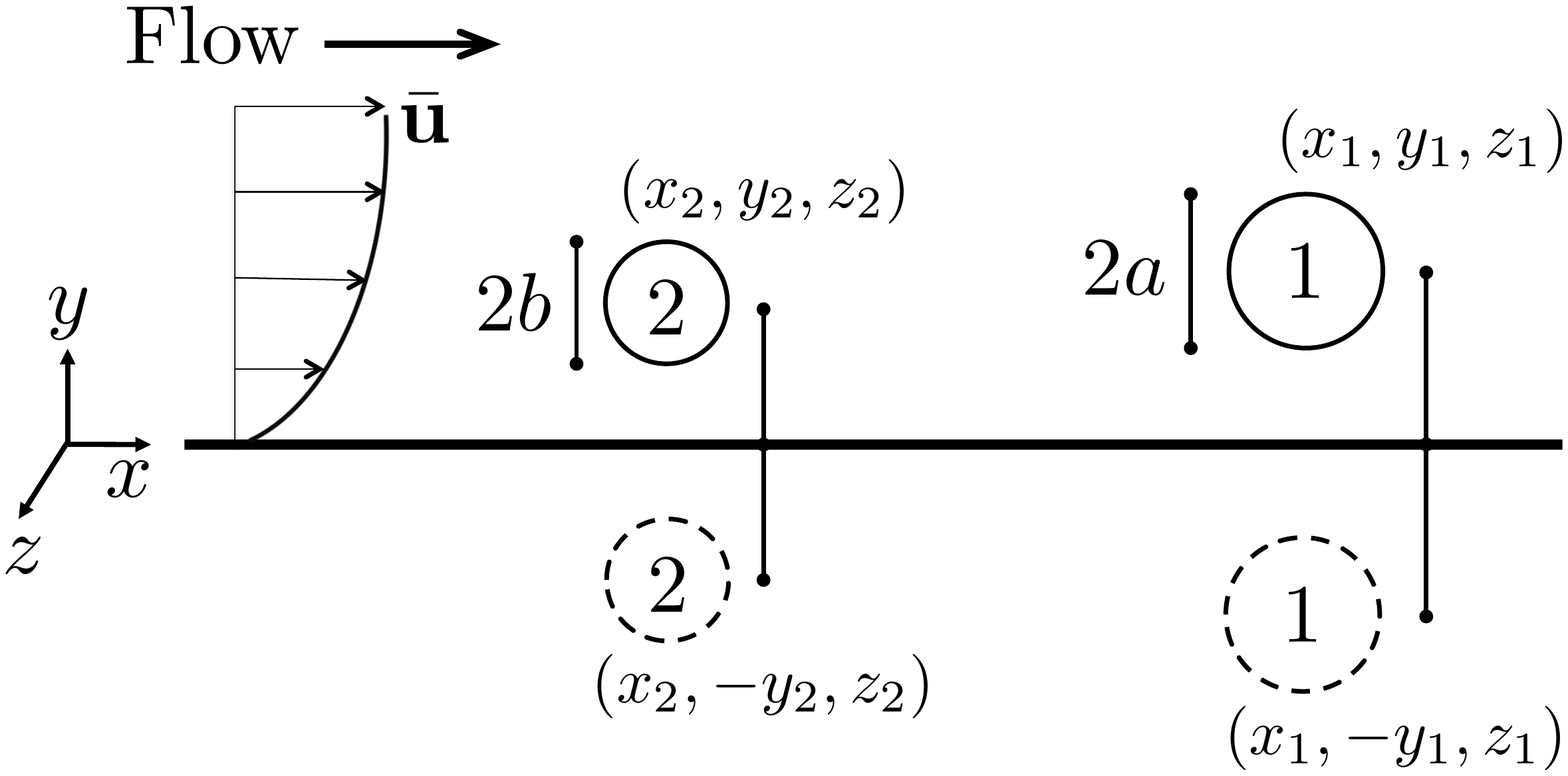}
\caption{Two particles near a wall at $y=0$ in a parabolic background flow $\ubar$. The particles have radius $a$ and $b$ respectively, and are located at positions $(x_1,y_1,z_1)$ and $(x_2,y_2,z_2)$. 
}\label{fig:supp_diagram}
\end{figure}

We consider the flow around two particles near a wall.  The particles have radius $a$ and $b$, respectively, and the wall is located at $y=0$.  The particles are centered at $(x_i,y_i,z_i)$ for $i=1,2$ and have image systems centered at the reflection point across the wall $(x_i,-y_i,z_i)$ (Figure \ref{fig:supp_diagram}).

We will derive this model for arbitrary parameters $a$, $b$, $x_i$, $y_i$, and $z_i$ for $i=1,2$ which satisfy the following assumptions:
\begin{equation}
      z_1 - z_2 \quad \ll \quad a,\,b \quad \ll \quad y_1,\, y_2 \quad \ll \quad x_1-x_2 \,.
\end{equation}
Once we have the model, in order to compare to experimental data, we will make the following substitutions:
\begin{equation}
      z_1 = z_2 = 0, \qquad a = b, \qquad y_1 = y_2 = h, \qquad \mbox{and} \quad dx = x_1-x_2.
\end{equation}
The objective is to find the separation length $dx$ for which the particles have no relative motion.

\subsection{Flow around a single particle in the background flow.}

The background flow in a rectangular channel is the Poisuille flow \cite{Papanastasiou99}.  Here we will approximate this flow by its Taylor expansion in the coordinate $y$,
\begin{equation}
      \ubar \sim \beta + \gamma_y (y-y_i) + \delta_{yy} (y-y_i)^2 \,.
\end{equation}

The flow around the particle can be derived using {Lamb's} solution for the flow exterior to a sphere \cite{Lamb45,KimKarrila2005}. Here we will only keep the term{s} that are $O(r^{-2})$ and $O(r^{-3})$.  In order to derive the image system in the next step, we must convert {Lamb's} solution into multipole singularities.  In this case the $O(r^{-2})$ term becomes the stresslet, and the $O(r^{-3})$ term is decomposed into the source dipole $\bv^{\mathrm{D}}$ and two stokeslet quadrupoles $\bv^{\mathrm{SQ}}$ and $\bw^{\mathrm{SQ}}$. Define $\br_i = (x-x_i,y-y_i,z-z_i)$ and $r_i = |\br_i|${. Then} the flow $\bv_i^0 \sim \bv_i^{\mathrm{ST}} +\bv_i^{\mathrm{D}} + \bv_i^{\mathrm{SQ}} + \bw_i^{\mathrm{SQ}} $ around each particle satisfies{\color{red}:}
\begin{align}\label{eq:stresslet}
      \bv_i^{\mathrm{ST}}  &= -\frac{5\gamma_y}{2} \left[ \frac{(x-x_i)(y-y_i)}{r_i^3} \br_i \right]\frac{1}{r_i^2} \,, \qquad 
      \bv_i^{\mathrm{D}} = -\frac{7\delta_{yy}}{24} \left[ \be_x - \frac{3(x-x_i)}{r_i^2}\br_i \right] \frac{1}{r_i^3} \,, \\
      \bv_i^{\mathrm{SQ}} &= -\frac{\dyy}{12} \bigg[ \be_x - \frac{3(y-y_i)^2}{r_i^2}\be_x - \frac{3(x-x_i)}{r_i^2}\br_i + \frac{15(x-x_i)(y-y_i)^2}{r_i^4}\br_i \bigg] \frac{1}{r_i^3} \,,\\
      \bw_i^{\mathrm{SQ}} &= -\frac{5\dyy}{24} \bigg[ - \be_x + \frac{3(y-y_i)^2}{r_i^2}\be_x -\frac{6(x-x_i)(y-y_i)}{r_i^2}\be_y - \frac{3(x-x_i)}{r_i^2}\br_i  + \frac{15(x-x_i)(y-y_i)^2}{r_i^4}\br_i  \bigg]\frac{1}{r_i^3} \,.  
\end{align}
 
\subsection{Image system for each particle due to the wall.}

Blake derived the image system for a stokeslet \cite{blake1971note}.  Using a similar procedure we derive the image systems for a stresslet and a source dipole. Define $\bigR_i = (x-x_i,y+y_i,z-z_i)$ and $R_i = |\bigR_i|$.

The image system for the stresslet is:
\begin{align}
      \bv_i^{\mathrm{STim}} &=  \bigg[ \frac{(x-x_i)(y+y_i)}{2R_i^3}\bigR_i - \frac{5 y y_i(x-x_i)(y+y_i)}{R_i^5}\bigR_i + \frac{y y_i(y+y_i)}{R_i^3}\be_x 
      - \frac{y_i^2(x-x_i)}{R_i^3}\be_y \bigg] \frac{5\gamma_y}{R_i^2} 
\end{align}

The image system for the source dipole is:
\begin{equation}
      \bv_i^{\mathrm{Dim}} = \frac{\delta_{yy}}{4} \left[ { \frac{30y(x-x_i)(y+y_i)}{R_i^4}\bigR_i - \frac{3(x-x_i)}{R_i^2}\bigR_i  - \frac{6y(y+y_i)}{R_i^2}\be_x + \frac{6y_i(x-x_i)}{R_i^2}\be_y - \be_x } \right] \frac{1}{R_i^3}
\end{equation}

The image system for the first stokeslet quadrupole is:
\begin{align}
      \bv_i^\mathrm{SQim} &= -\frac{\dyy}{24} \bigg[ \frac{3(x-x_i)}{R_i^2}\bigR_i - \frac{15(x-x_i)(y+y_i)^2}{R_i^4}\bigR_i - \frac{30y y_i(x-x_i)}{R_i^4}\bigR_i + \frac{210y y_i (x-x_i)(y+y_i)^2}{R_i^6}\bigR_i  \\ &\phantom{{}=\hspace{.05in}} - \frac{30y_i(x-x_i)(y^2-y_i^2)}{R_i^4}\be_y  - \frac{6y_i(x-x_i)}{R_i^2}\be_y - \frac{30y y_i (y+y_i)^2}{R_i^4}\be_x  + \frac{3(y+y_i)^2}{R_i^2}\be_x  + \frac{6y y_i}{R_i^2}\be_x  -\be_x \bigg]\frac{1}{R_i^3} \,. \notag
\end{align}

The image system for the second stokeslet quadrupole is:
\begin{align}
      \bw_i^{\mathrm{SQim}} &= -\frac{5\dyy}{24} \bigg[ \frac{3(x-x_i)}{R_i^2}\bigR_i - \frac{15(x-x_i)(y+y_i)^2}{R_i^4}\bigR_i - \frac{30{y y_i}(x-x_i)}{R_i^4}\bigR_i + \frac{210 y y_i (x-x_i) (y+y_i)^2}{R_i^6}\bigR_i\\  &\phantom{{}=\hspace{.5in}} { - \frac{60y(x-x_i)(y+y_i)}{R_i^4}\bigR_i}    -\frac{30y_i(x-x_i)(y^2-y_i^2)}{R_i^4}\be_y  + \frac{6(x-x_i)(y-2y_i)}{R_i^2}\be_y \notag \\ &\phantom{{}=\hspace{.5in}}   {-} \frac{{\color{red}4}y(2y+3y_i)}{R_i^2}\be_x  - \frac{3(y+y_i)^2}{R_i^2}\be_x  - \frac{{10 y^2 (x-x_i)^2}}{R_i^4}\be_x + \be_x {  + \frac{10y y_i (z-z_i)^2}{R_i^4}\be_x } \notag \\ &\phantom{{}=\hspace{.5in}} {+ \frac{20y(y+y_i)(z-z_i)^2}{R_i^4}\be_x + \frac{30y(y+y_i)(x-x_i)^2}{R_i^4}\be_x + \frac{20y^2(y+y_i)^2}{R_i^4}\be_x  } \bigg]\frac{1}{R_i^2} \,. \notag
\end{align}

\subsection{Induced velocities from particle and image system.}

Up to this point we have derived the flow around each particle due to the background flow and the wall,  
\begin{equation}\label{eq:supp_vel_1p}
      \bv_i \sim \,  (\bv_i^{\mathrm{ST}} + \bv_i^{\mathrm{STim}}) + (\bv_i^{\mathrm{D}}  + \bv_i^{\mathrm{Dim}}) + (\bv_i^{\mathrm{SQ}} + \bv_i^{\mathrm{SQim}}) + (\bw_i^{\mathrm{SQ}} + \bw_i^{\mathrm{SQim}}) \,.
\end{equation}
For each particle we can compute the induced velocity from the other particle and image system,
\begin{eqnarray}
      \bU_1 = \left( 1 + \frac{a^2}{6} \nabla^2 \right) (\ubar + \bv_2) \Big|_{\bx = \bx_1} \, , \qquad
       \bU_2 = \left( 1 +  \frac{b^2}{6} \nabla^2 \right) (\ubar + \bv_1) \Big|_{\bx = \bx_2} \,.
\end{eqnarray}
Then we define the relative velocity $d\bU = \bU_2 - \bU_1$.  The particles are considered to be in equilibrium when their induced velocities are equal, that is $d\bU = 0$.

{

\subsection{Second order system of ODEs}

Here, we make the assumption that $y_1$ and $y_2$ are symmetric about the inertial focusing line at $y=h$.  That is, 
\begin{equation}
y_1 = h- \frac{dy}{2}\,, \quad \mbox{and} \quad y_2 = h+ \frac{dy}{2}\,.
\end{equation}
Substituting this into Equations (9)-(10) in the main text, we arrive at:
\begin{align}\label{eq:ode2_dx}
      \dot{dx} &= dU, \, \qquad \qquad \qquad   dx(t=0) = k_0d, \\ \label{eq:ode2_dy}
      \dot{dy} &= dV - \Gamma dy , \qquad \quad dy(t=0) = 0. 
\end{align}
Then we can write out the forms for $dU$ and $dV$. Here, we define $r^2 = dx^2 + dy^2$ and $R^2 = dx^2 + 4h^2$.

\begin{align}
dU = &-dy \gy +\frac{20 a^5 dy \gy dx^{10}}{3 r^7 R^8}
	-\frac{20 a^5 dy \gy dx^{10}}{3 r^6 R^9}
	-\frac{6 a^5 dy h \dyy dx^{10}}{r^6 R^9}
	-\frac{5 a^5 dy^3 \gy dx^8}{3 r^7 R^8}
	+\frac{320 a^5 dy h^2 \gy dx^8}{3 r^7 R^8} \notag \\
	&-\frac{20 a^5 dy^3 \gy dx^8}{r^6 R^9}
	+\frac{180 a^5 dy h^2 \gy dx^8}{r^6 R^9}
	-\frac{18 a^5 dy h^3 \dyy dx^8}{r^6 R^9}
	-\frac{18 a^5 dy^3 h \dyy dx^8}{r^6 R^9}
	+\frac{640 a^5 dy h^4 \gy dx^6}{r^7 R^8} \notag \\
	&-\frac{80 a^5 dy^3 h^2 \gy dx^6}{3 r^7 R^8}
	-\frac{20 a^5 dy^5 \gy dx^6}{r^6 R^9}
	-\frac{320 a^5 dy h^4 \gy dx^6}{3 r^6 R^9}
	+\frac{540 a^5 dy^3 h^2 \gy dx^6}{r^6 R^9}
	+\frac{24 a^5 dy h^5 \dyy dx^6}{r^6 R^9} \notag \\
	&-\frac{54 a^5 dy^3 h^3 \dyy dx^6}{r^6 R^9}
	-\frac{18 a^5 dy^5 h \dyy dx^6}{r^6 R^9}
	+\frac{5120 a^5 dy h^6 \gy dx^4}{3 r^7 R^8}
	-\frac{160 a^5 dy^3 h^4 \gy dx^4}{r^7 R^8}
	-\frac{20 a^5 dy^7 \gy dx^4}{3 r^6 R^9}\\
	&-\frac{320 a^5 dy^3 h^4 \gy dx^4}{r^6 R^9}
	+\frac{540 a^5 dy^5 h^2 \gy dx^4}{r^6 R^9}
	+\frac{72 a^5 dy^3 h^5 \dyy dx^4}{r^6 R^9}
	-\frac{54 a^5 dy^5 h^3 \dyy dx^4}{r^6 R^9}
	-\frac{6 a^5 dy^7 h \dyy dx^4}{r^6 R^9} \notag \\
	&-\frac{105 a^7 dy h \dyy dx^4}{R^{11}}
	-\frac{5 a^3 dy \gy dx^2}{r^5}
	+\frac{5120 a^5 dy h^8 \gy dx^2}{3 r^7 R^8}
	-\frac{1280 a^5 dy^3 h^6 \gy dx^2}{3 r^7 R^8}
	-\frac{320 a^5 dy^5 h^4 \gy dx^2}{r^6 R^9} \notag \\
	&+\frac{180 a^5 dy^7 h^2 \gy dx^2}{r^6 R^9}
	+\frac{72 a^5 dy^5 h^5 \dyy dx^2}{r^6 R^9}
	-\frac{18 a^5 dy^7 h^3 \dyy dx^2}{r^6 R^9}
	+\frac{2870 a^7 dy h^3 \dyy dx^2}{3 R^{11}}	
	-\frac{1280 a^5 dy^3 h^8 \gy}{3 r^7 R^8} \notag \\
	&-\frac{320 a^5 dy^7 h^4 \gy}{3 r^6 R^9}
	+\frac{24 a^5 dy^7 h^5 \dyy}{r^6 R^9}
	-\frac{1120 a^7 dy h^5 \dyy}{3 R^{11}} \notag 
\end{align}

\begin{align}
dV = &-\frac{5 a^5 \gy dx^{11}}{3 r^7 R^8}
	-\frac{5 a^5 \gy dx^{11}}{3 r^6 R^9}
	+\frac{20 a^5 dy^2 \gy dx^9}{3 r^7 R^8}
	-\frac{80 a^5 h^2 \gy dx^9}{3 r^7 R^8}
	-\frac{5 a^5 dy^2 \gy dx^9}{r^6 R^9}
	+\frac{120 a^5 h^2 \gy dx^9}{r^6 R^9}
	-\frac{135 a^5 h^3\dyy dx^9}{r^6 R^9} \notag \\
	&+\frac{105 a^5 dy^2 h \dyy dx^9}{4 r^6 R^9}
	-\frac{5 a^3 dy^2 \gy dx^7}{r^5 R^6}
	+\frac{5 a^3 dy^2 \gy dx^7}{2 r^4 R^7}
	-\frac{10 a^3 h^2 \gy dx^7}{r^4 R^7}
	-\frac{160 a^5 h^4 \gy dx^7}{r^7 R^8}
	+\frac{320 a^5 dy^2 h^2 \gy dx^7}{3 r^7 R^8} \notag \\
	&-\frac{5 a^5 dy^4 \gy dx^7}{r^6 R^9}
	-\frac{1280 a^5 h^4 \gy dx^7}{3 r^6 R^9}
	+\frac{360 a^5 dy^2 h^2 \gy dx^7}{r^6 R^9}
	+\frac{440 a^5 h^5 \dyy dx^7}{r^6 R^9}
	-\frac{545 a^5 dy^2 h^3 \dyy dx^7}{r^6 R^9} \notag \\
	&+\frac{315 a^5 dy^4 h \dyy dx^7}{4 r^6 R^9}
	-\frac{60 a^3 dy^2 h^2 \gy dx^5}{r^5 R^6}
	+\frac{5 a^3 dy^4 \gy dx^5}{r^4 R^7}
	+\frac{160 a^3 h^4 \gy dx^5}{r^4 R^7}
	-\frac{60 a^3 dy^2 h^2 \gy dx^5}{r^4 R^7}
	-\frac{1280 a^5 h^6 \gy dx^5}{3 r^7 R^8}  \notag \\
	&+\frac{640 a^5 dy^2 h^4 \gy dx^5}{r^7 R^8}
	-\frac{5 a^5 dy^6 \gy dx^5}{3 r^6 R^9}
	-\frac{1280 a^5 dy^2 h^4 \gy dx^5}{r^6 R^9}
	+\frac{360 a^5 dy^4 h^2 \gy dx^5}{r^6 R^9}
	+\frac{1320 a^5 dy^2 h^5 \dyy dx^5}{r^6 R^9} \notag \\
	&-\frac{825 a^5 dy^4 h^3 \dyy dx^5}{r^6 R^9}
	+\frac{315 a^5 dy^6 h \dyy dx^5}{4 r^6 R^9}
	-\frac{50 a^7 h \dyy dx^5}{R^{11}}
	-\frac{240 a^3 dy^2 h^4 \gy dx^3}{r^5 R^6}
	+\frac{5 a^3 dy^6 \gy dx^3}{2 r^4 R^7}\\
	&+\frac{320 a^3 dy^2 h^4 \gy dx^3}{r^4 R^7}
	-\frac{90 a^3 dy^4 h^2 \gy dx^3}{r^4 R^7}
	-\frac{1280 a^5 h^8 \gy dx^3}{3 r^7 R^8}
	+\frac{5120 a^5 dy^2 h^6 \gy dx^3}{3 r^7 R^8}
	-\frac{1280 a^5 dy^4 h^4 \gy dx^3}{r^6 R^9} \notag \\
	&+\frac{120 a^5 dy^6 h^2 \gy dx^3}{r^6 R^9}
	+\frac{1320 a^5 dy^4 h^5 \dyy dx^3}{r^6 R^9}
	-\frac{555 a^5 dy^6 h^3 \dyy dx^3}{r^6 R^9}
	+\frac{105 a^5 dy^8 h \dyy dx^3}{4 r^6 R^9}
	+\frac{2650 a^7 h^3 \dyy dx^3}{3 R^{11}} \notag \\
	&-\frac{320 a^3 dy^2 h^6 \gy dx}{r^5 R^6}
	+\frac{160 a^3 dy^4 h^4 \gy dx}{r^4 R^7}
	-\frac{40 a^3 dy^6 h^2 \gy dx}{r^4 R^7}
	+\frac{5120 a^5 dy^2 h^8 \gy dx}{3 r^7 R^8}
	-\frac{1280 a^5 dy^6 h^4 \gy dx}{3 r^6 R^9} \notag \\
	&+\frac{440 a^5 dy^6 h^5 \dyy dx}{r^6 R^9}
	-\frac{140 a^5 dy^8 h^3 \dyy dx}{r^6 R^9}
	-\frac{4640 a^7 h^5 \dyy dx}{3 R^{11}} \notag 
\end{align}

Consider the case used in the main text and in Kahkeshani \etal \cite{kahkeshani2016preferred}, where $AR = 1.7$, $W = 60\mu$m, $H=35\mu$m, and $a = 6\mu$m.  We can calculate $h$ from Hood \etal \cite{Hood15} and expect that $h = 11.6\mu$m.  We can also calculate $\gy = 2.4\mu$m/s and $\dyy = -0.2\mu$m/s$^2$. At $\Rey=1$, the inertial constant is $\Gamma = -0.0315$. Plugging this into the system of ODEs in Equations (\ref{eq:ode2_dx})-(\ref{eq:ode2_dy}), we find:
\begin{align}\label{eq:ode_dx_kahk}
      \dot{dx} &= dU^*, \,    \\ \label{eq:ode_dy_kahk}
      \dot{dy} &= dV^* - \Gamma dy ,  \\ \label{eq:ode_init_kahk}
      dx(0) &= k_0d, \qquad \quad dy(0) = 0.
\end{align}
where
\begin{align}\label{eq:du_kahk}
dU^* = &-2.436 dy + \frac{126282. dx^{10} dy}{r^7 R^8}
	-\frac{258.424 dx^{10} dy}{r^6 R^9}
	-\frac{31570.6 dx^8 dy^3}{r^7 R^8}
	-\frac{775.272 dx^8 dy^3}{r^6 R^9}
	+\frac{2.71881\times 10^8 dx^8 dy}{r^7 R^8} \notag \\
	&+\frac{5.09672\times 10^8 dx^8 dy}{r^6 R^9}
	-\frac{775.272 dx^6 dy^5}{r^6 R^9}
	-\frac{6.79702\times 10^7 dx^6 dy^3}{r^7 R^8}
	+\frac{1.52902\times 10^9 dx^6 dy^3}{r^6 R^9} \notag \\
	&+\frac{2.19506\times 10^{11} dx^6 dy}{r^7 R^8}
	-\frac{4.57116\times 10^{10} dx^6 dy}{r^6 R^9}
	-\frac{258.424 dx^4 dy^7}{r^6 R^9}
	+\frac{1.52902\times 10^9 dx^4 dy^5}{r^6 R^9}\\
	&-\frac{5.48764\times 10^{10} dx^4 dy^3}{r^7 R^8}
	-\frac{1.37135\times 10^{11} dx^4 dy^3}{r^6 R^9}
	+\frac{7.87644\times 10^{13} dx^4 dy}{r^7 R^8}
	+\frac{7.9395\times 10^7 dx^4 dy}{R^{11}} \notag \\
	&+\frac{5.09672\times 10^8 dx^2 dy^7}{r^6 R^9}
	-\frac{1.37135\times 10^{11} dx^2 dy^5}{r^6 R^9}
	-\frac{1.96911\times 10^{13} dx^2 dy^3}{r^7 R^8}
	+\frac{1.05985\times 10^{16} dx^2 dy}{r^7 R^8} \notag \\
	&-\frac{2630.88 dx^2 dy}{r^5}
	-\frac{9.73376\times 10^{10} dx^2 dy}{R^{11}}
	-\frac{4.57116\times 10^{10} dy^7}{r^6 R^9}
	-\frac{2.64964\times 10^{15} dy^3}{r^7 R^8}
	+\frac{5.11131\times 10^{12} dy}{R^{11}} \,,\notag 
\end{align}
and
\begin{align}\label{eq:dv_kahk}
dV^* = &-\frac{31570.6 dx^{11}}{r^7 R^8}
	-\frac{31570.6 dx^{11}}{r^6 R^9}+\frac{126282. dx^9 dy^2}{r^7 R^8}
	-\frac{646066. dx^9 dy^2}{r^6 R^9}
	-\frac{6.79702\times 10^7 dx^9}{r^7 R^8}
	+\frac{6.87415\times 10^8 dx^9}{r^6 R^9} \notag \\
	&-\frac{1.74877\times 10^6 dx^7 dy^4}{r^6 R^9}
	+\frac{2.71881\times 10^8 dx^7 dy^2}{r^7 R^8}
	+\frac{2.45793\times 10^9 dx^7 dy^2}{r^6 R^9}
	-\frac{2630.88 dx^7 dy^2}{r^5 R^6}
	+\frac{1315.44 dx^7 dy^2}{r^4 R^7} \notag \\
	&-\frac{5.48764\times 10^{10} dx^7}{r^7 R^8}
	-\frac{3.13672\times 10^{11} dx^7}{r^6 R^9}
	-\frac{708022. dx^7}{r^4 R^7}
	-\frac{1.68563\times 10^6 dx^5 dy^6}{r^6 R^9}
	+\frac{3.24929\times 10^9 dx^5 dy^4}{r^6 R^9} \notag \\
	&+\frac{2630.88 dx^5 dy^4}{r^4 R^7}
	+\frac{2.19506\times 10^{11} dx^5 dy^2}{r^7 R^8}
	-\frac{9.41015\times 10^{11} dx^5 dy^2}{r^6 R^9}
	-\frac{4.24813\times 10^6 dx^5 dy^2}{r^5 R^6}\\
	&-\frac{4.24813\times 10^6 dx^5 dy^2}{r^4 R^7}
	-\frac{1.96911\times 10^{13} dx^5}{r^7 R^8}
	+\frac{1.52434\times 10^9 dx^5}{r^4 R^7}
	+\frac{3.78071\times 10^7 dx^5}{R^{11}}
	-\frac{551354. dx^3 dy^8}{r^6 R^9} \notag \\
	&+\frac{1.87446\times 10^9 dx^3 dy^6}{r^6 R^9}
	+\frac{1315.44 dx^3 dy^6}{r^4 R^7}
	-\frac{9.41015\times 10^{11} dx^3 dy^4}{r^6 R^9}
	-\frac{6.3722\times 10^6 dx^3 dy^4}{r^4 R^7} \notag \\
	&+\frac{7.87644\times 10^{13} dx^3 dy^2}{r^7 R^8}
	-\frac{2.28652\times 10^9 dx^3 dy^2}{r^5 R^6}
	+\frac{3.04869\times 10^9 dx^3 dy^2}{r^4 R^7}
	-\frac{2.64964\times 10^{15} dx^3}{r^7 R^8} \notag \\
	&-\frac{8.98762\times 10^{10} dx^3}{R^{11}}
	+\frac{3.95681\times 10^8 dx dy^8}{r^6 R^9}
	-\frac{3.13672\times 10^{11} dx dy^6}{r^6 R^9}
	-\frac{2.83209\times 10^6 dx dy^6}{r^4 R^7} \notag \\
	&+\frac{1.52434\times 10^9 dx dy^4}{r^4 R^7}
	+\frac{1.05985\times 10^{16} dx dy^2}{r^7 R^8}
	-\frac{4.10231\times 10^{11} dx dy^2}{r^5 R^6}
	+\frac{2.11754\times 10^{13} dx}{R^{11}} \notag 
\end{align}


}

\subsection{The separation length for two particles.}

Now we want to find the separation length for two particles.  We are going to assume that the particles are the same size and are already inertially focused, that is:
\begin{equation}
      a = b, \qquad y_1 = y_2 = h, \qquad z_1 = z_2 = 0 \,.
\end{equation}
Then we define the separation length as $dx = x_1 - x_2$.  The objective is to find $dx$ for which $d\bU = 0$. 

Define $r^2 = (x_1-x_2)^2 + (y_1-y_2)^2 +(z_1-z_2)^2 = dx^2$ and $R^2 = (x_1-x_2)^2 + (y_1+y_2)^2 +(z_1-z_2)^2 = dx^2 + 4h^2${.} Let $d\bU = (dU, dV, dW)${. Then} each component satisfies:
\begin{align}
      dU &= -\frac{10\gy h dx^2 (dx^2 + 5h^2) }{3 (dx^2 + 4h^2)^{7/2}} \,, \\
      dV &= -\frac{5 \gy a^2 }{3 dx^4} +\frac{10h^2\gy dx(dx^4 - 12 h^2 dx^2  - 64 h^4)}{(dx^2 + 4h^2)^{9/2}}  + \frac{5a^2\gy dx(dx^4 - 72 h^2 dx^2 + 256 h^4)}{3(dx^2 + 4h^2)^{9/2}} \\
      &+ \frac{5h^3\dyy dx(27 dx^4 + 20 h^2 dx^2 - 352 h^4)}{(dx^2 + 4h^2)^{11/2}} + \frac{10h  a^2\dyy dx (15 dx^4 - 265 h^2 dx^2 + 464 h^4)}{3(dx^2 + 4h^2)^{11/2}}  \notag  \,, \\
      dW &= 0 \,.
\end{align}

Now we can make an ODE for $dx$ and $dy$.  Additionally we will include the inertial lift velocity which acts on $dy$.
\begin{align}
      \dot{dx} &\sim dU \\ 
      \dot{dy} &\sim dV + \Gamma dy 
\end{align}
Now using the experimental values for $h$ and $a$, along with $\gy$, $\dyy$, and $\Gamma$, we can solve this system of ODEs numerically. 

\begin{figure}[h]
\centering
\includegraphics[scale=.5]{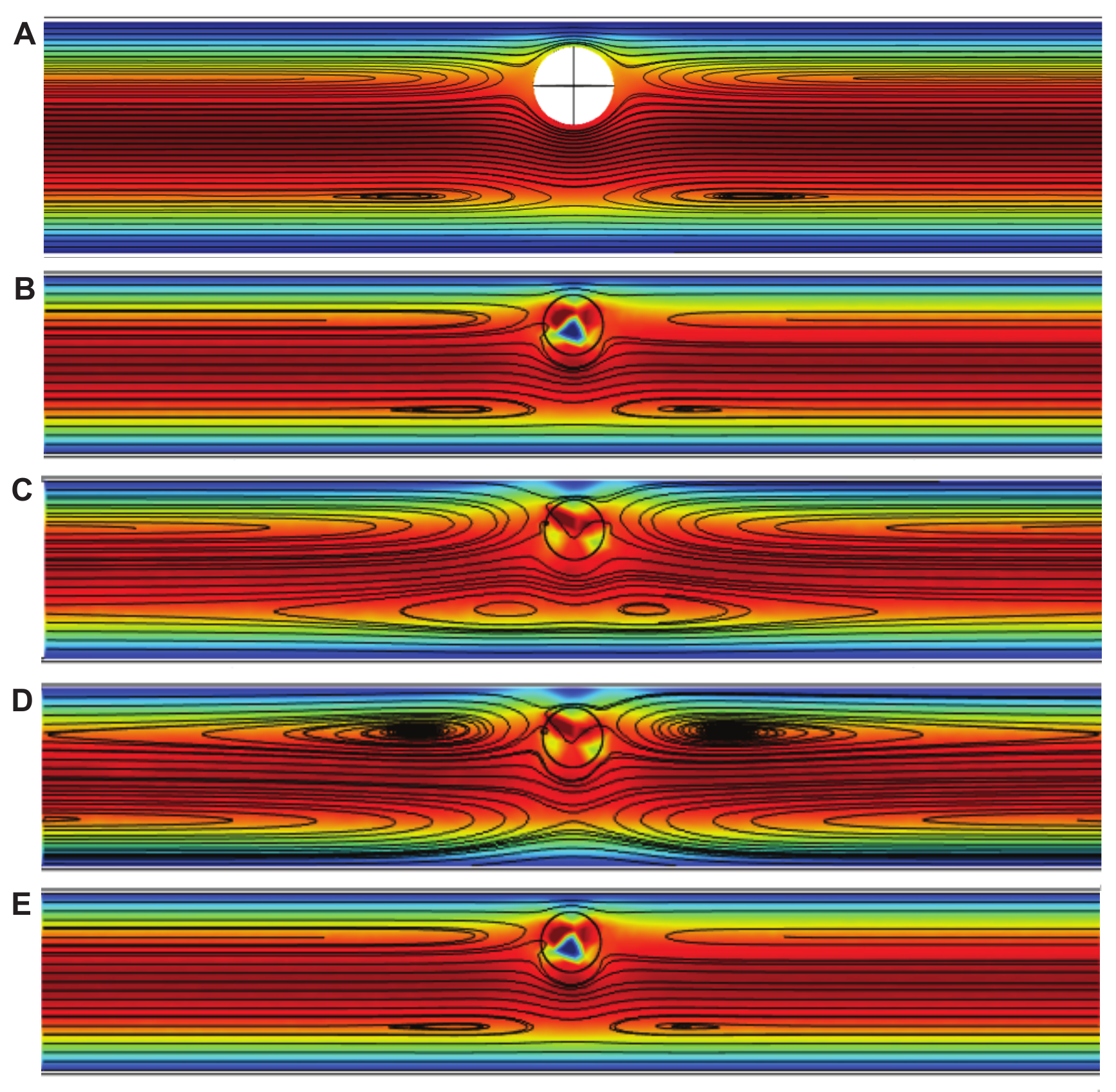}
\caption{Fluid streamlines around a single particle in its equilibrium position for (A) the full NSE solution $\bu$, (B) the stresslet $\bv^{ST}$ (C) the stresslet plus the first image $\bv^{ST} + \bv_1$ (D) the stresslet plus two images $\bv^{ST} + \bv_1+\bv_2$, and (E) the stresslet plus the full computational image $\bv^{ST} + \bv_{\infty}$.}\label{fig:images}
\end{figure}

\section{Cross-stream eddies in the asymptotic model}

Here we show the cross-stream eddies appear in the asymptotic expansion of the flow around a single inertially-focused particle.  We consider a channel with dimensions: $H=35\mu$m, $W = 60\mu$m, $AR = 1.7$, and $\alpha = 0.17$.

Let $\bu$ solve the full NSE with exact boundary conditions. We solve for $\bu$ numerically using finite element methods in Comsol Multiphysics (Los Angeles, CA).  We validate this numerical method in Section \ref{sec:validate}.

In the channel, the closest wall to the inertially focused particle is at $y = +H/2$. Recall that $\bv^{ST}$ is the stresslet, shown in equation (\ref{eq:stresslet}).  Let $\bv_1$, $\bv_2$ and $\bv_{\infty}$ all solve the Stokes equations and {satisfy} the following boundary conditions:
\begin{align}
 \bv_1 &= -\bv^{ST} \mbox{ on } y = +H/2\,,  {\qquad \bv_1 = 0 \mbox{ on remaining walls }} \\
 \bv_2 &= -\bv_1 \mbox{ on } y = -H/2\,, {\qquad \bv_2 = 0 \mbox{ on remaining walls }} \\
 \bv_{\infty} &= -\bv^{ST} \mbox{ on all the channel walls.}
\end{align}
{ All the $\bv_i$ satisfy $\bv_i = 0$ at the inlet and outlet.} Note that 
\begin{equation}
 \bu = \bv^{ST} + \bv_{\infty} + \hdots, \qquad \mbox{and} \qquad \bv_{\infty} \sim \bv_1 + \bv_2 + \hdots\,.
\end{equation}

We plot the streamlines for $\bu$, $\bv^{ST}$, $\bv_1$, $\bv_2$, and $\bv_{\infty}$ in Figure \ref{fig:images}.  Note that the cross stream eddies in $\bu$ only appear in $\bv^{ST} + \bv_1$ and $\bv^{ST} + \bv_{\infty}$ (Figure \ref{fig:images} A,C,E). The cross-stream eddies appear after reflecting across the wall at $y = -H/2$.  By construction, we have demonstrated that the cross-stream eddies can be quantitatively reproduced using the same model we develop for same streamline interactions, i.e. viscous particle interactions with the wall.

\begin{figure}[h]
\centering
\includegraphics[scale=.5]{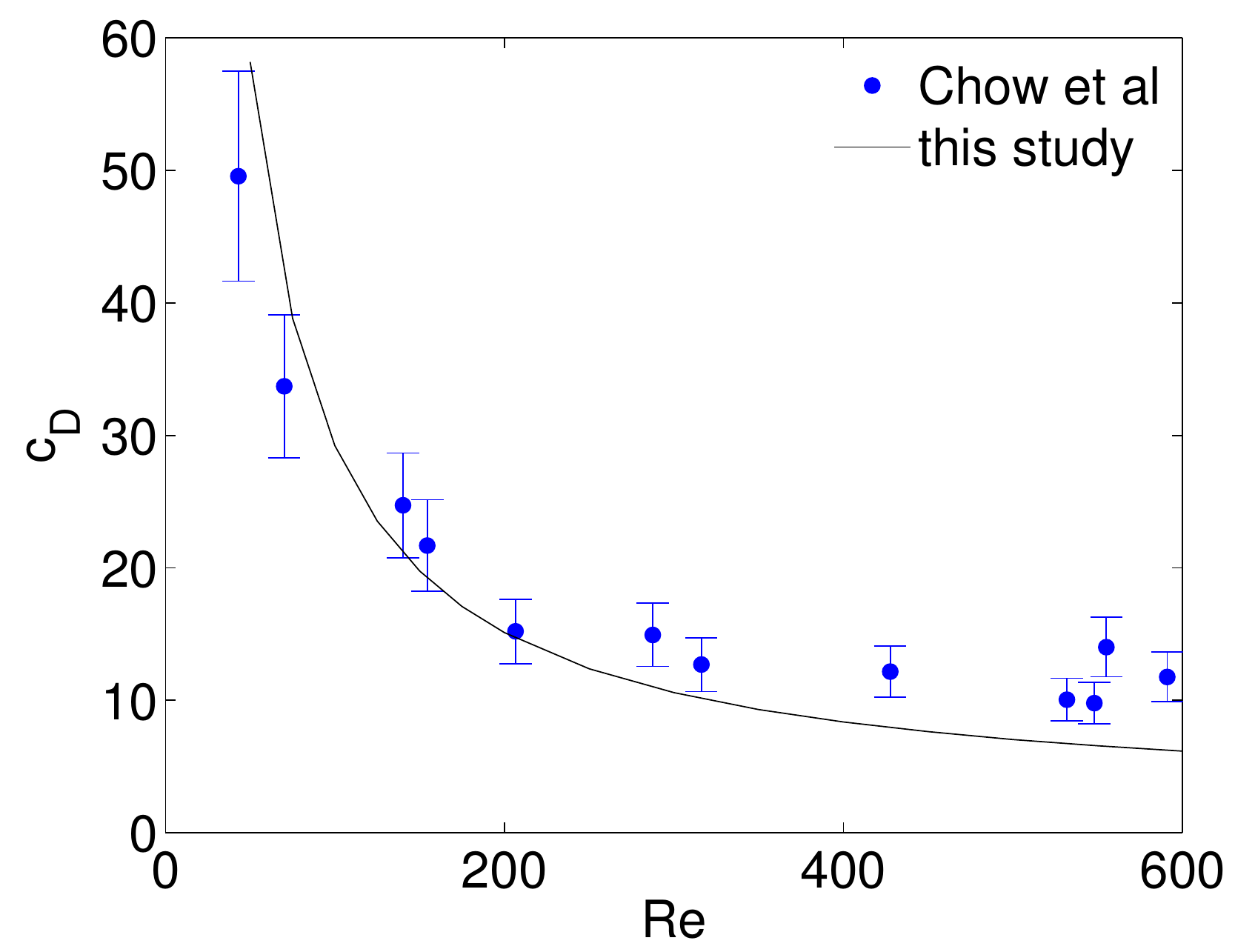}
\caption{\label{fig:drag}Drag coefficient dependence on Reynolds number.  The results from our numerical solver compare well with the data from Chow \etal \cite{chow1989drag}}
\end{figure}

\section{Validating the numerical solver against data}\label{sec:validate}

To test the accuracy of our numerical solver, we compare to experimental measurements of the drag coefficient of a sphere in a square channel.  Chow \etal observed the drag coefficient of a sphere whose diameter $d$ is very close to the width $W$ of a square channel \cite{chow1989drag}.  For this comparision, we use their measurements for the size ratio $d/W = 0.886 \pm 0.008$.  The Reynolds number of the flow was defined by $\Rey = UW/\nu$, where $U$ is the average fluid velocity in an empty channel and $\nu$ is the kinematic viscosity of water. Let $\rho$ denote the density of water. 

Our numerical solver modeled a square channel with lengths scaled by the channel width $W$.  That is, the square channel had dimensions $1 \times 1 \times 6$, and the particle had radius $a = 0.443$.  We used Comsol Multiphysics (Los Angles, CA) to solve the PDE with variable Reynolds numbers.  The particle velocity $U_p$ was chosen arbitrarily to be $U_P = .75U$.  We measured the drag force $F_D$ on the particle using Lagrange multipliers.

{ According to the drag equation, the drag on a sphere satisfies:
\begin{equation}\label{eq:fd}
F_d = \frac{1}{2} C_D \rho A V^2 \,,
\end{equation}
Where $A$ is the projected area of the sphere satisfies $A = \pi a^2$, and $V$ is the speed of the object relative to the fluid, specifically $V = U_p - U$. Rearranging the terms in equation (\ref{eq:fd}), we arrive at a formula for the drag coefficient:}

\begin{equation}\label{eq:drag}
      C_D = \frac{2F_D}  {\pi a^2 \rho (U_p - U)^2} .
\end{equation}
{ Note that $U_p$ was chosen to avoid division by zero in equation (\ref{eq:drag}). Any choice of $U_p$ that satisfies $U_p - U \gg 0$ should suffice. }

The results from our numerical solver compare well with the data from Chow \etal \cite{chow1989drag}, especially for $\Rey \le 300$ (Figure \ref{fig:drag}).  The range of Reynolds numbers from experiments is $30 \le \Rey \le 110$, which is well within the numerical range of accuracy.


\providecommand{\noopsort}[1]{}\providecommand{\singleletter}[1]{#1}%

\end{document}